\documentclass[aps,onecolumn,showpacs,showkeys,nofootinbib]{revtex4}
\usepackage{epsfig}
\usepackage{amsmath}
\usepackage{amsfonts}
\usepackage{amssymb}
\usepackage{graphicx}
\usepackage{colordvi}
\usepackage{rotating}

\begin{document}

\title{Casimir effect in Extended Theories of Gravity}
\author{$^{1,2}$G. Lambiase\footnote{lambiase@sa.infn.it}, $^3$A. Stabile\footnote{arturo.stabile@unisannio.it}, $^{1,2}$An. Stabile\footnote{anstabile@gmail.com}}

\affiliation{$^1$Dipartimento di Fisica ``E.R. Caianiello'', Universit\`{a} degli Studi di Salerno, via G. Paolo II, Stecca 9, I
 - 84084 Fisciano, Italy\\
$^2$Istituto Nazionale di   Fisica Nucleare (INFN) Sezione di Napoli, Gruppo collegato di Salerno\\
$^3$Dipartimento di Ingegneria, Universit\`a degli Studi del Sannio, 
Palazzo Dell'Aquila Bosco Lucarelli, Corso Garibaldi, 107
- 82100, Benevento, Italy}

\begin{abstract}
We study the Casimir vacuum energy density and the Casimir pressure for a massless scalar field confined between two nearby parallel plates in a slightly curved, static spacetime background, employing the weak field approximation in the framework of Extended Theories of Gravity (ETG). Following a perturbative approach upto second order, we find the gravity correction in the ETG to Casimir vacuum energy density and pressure. The corrections to the vacuum energy density in presence of curved spacetime in the framework of General Relativity (GR) are small and today they are still undetected with the current technology. However, future sensitivity improvement in gravitational interferometer experiments will give an useful  tool to detect such effect induced by gravity. For these reason we retain interesting from a theoretical point of view generalize the outcomes of GR in the context of ETG. Finally, we find the general relation to constraining the free parameters of the ETG.
\end{abstract}

\pacs{XXXX}
\maketitle

\section{Introduction}
Our Universe appears spatially flat and undergoing a period of accelerated expansion. Several observational data probe this picture~\cite{riess, ast, clo, spe,carrol,sahini}  but two unrevealed ingredients are needed in order to achieve this dynamical scenario, namely {\sl dark matter} at galactic and extragalactic scales and {\sl dark energy} at cosmological scales. The dynamical evolution of self-gravitating structures can be explained within Newtonian gravity, but a dark matter component is required in order to obtain agreement with observations~\cite{NFW}.

Lately, {\sl models of extended gravity}~\cite{repsergei,olmo_palatini} have been considered as a viable theoretical mechanism to explain cosmic acceleration and galactic rotation curves. In such models one extends only the geometric sector, without introducing any exotic matter. Such models may result from an effective theory of a quantum gravity formulation, which may contain additional contributions with respect to GR, at galactic, extra-galactic and  cosmological scales where, otherwise, large amounts of unknown dark components are required. In the context of {\it models of extended gravity},
one may consider that gravitational interaction acts differently  at different scales, whilst the robust results of GR at local and solar system scales are preserved \cite{book}.

In the so-called weak-field approximation, any relativistic theory of gravitation  yields, in general, corrections to the gravitational potentials ({\em e.g.}, Ref.~\cite{Qua91}) which, at the  Newtonian and post-Newtonian level, could constitute the test-bed for these theories \cite{Will93}. In fact in ETG there are further gravitational degrees of freedom and moreover the form of the gravitational interaction is no longer scale-invariant. Hence, in a given situation, besides the Schwarzschild radius, other characteristic gravitational scales could come out from dynamics. Such scales, in the weak field approximation, should exhibit a form of gravitational confinement in this way~\cite{annalen}.

Models of fourth order gravity have been studied mainly in the Newtonian limit (weak-field and small velocity) \cite{PRD1,mio2}, as well as in the Minkowskian limit~\cite{minko}. In the former one finds  modifications of the gravitational potential, whilst in the latter one obtains massive gravitational wave modes~\cite{quadrupolo,Sta1}. The weak-field limit of such proposals have to be tested against realistic self-gravitating systems. Galactic rotation curves, stellar hydrodynamics and gravitational lensing appear natural candidates as test-bed experiments~\cite{ BHL, BHL1, stabile_scelza,stabile_scelza2,stabstab,stabile_stabile_cap}.

These corrections to the gravitational Lagrangian were already considered by several authors~\cite{weyl_2, edd, lan,pauli,bach,buc,bic,FOG_CGL,FOG_CGL2,FOG_CGL3,FOG_CGL4,FOG_CGL5,FOG_CGL6,FOG_CGL7,FOG_CGL8}). From a conceptual viewpoint, there is no reason \emph{a priori} to restrict the gravitational Lagrangian to a linear function of the Ricci scalar minimally coupled to matter~\cite{mag-fer-fra}.  In particular, one may consider the generalization of $f({\cal R})$ models, where ${\cal R}$ is the Ricci scalar, through generic functions containing curvature invariants such as the \emph{Ricci squared} (${\cal R}_{\alpha\beta}{\cal R}^{\alpha\beta}$) or the \emph{Riemann squared} (${\cal R}_{\alpha\beta\gamma\delta}{\cal R}^{\alpha\beta\gamma\delta}$), which however are not independed  due to the Gauss-Bonnet invariant ${\cal R}^2-4{\cal R}_{\mu\nu}{\cal R}^{\mu\nu}+{\cal R}_{\mu\nu\lambda\sigma}
{\cal R}^{\mu\nu\lambda\sigma}$. Note that the same remark applies to the Weyl invariant $C_{\alpha\beta\gamma\delta}C^{\alpha\beta\gamma\delta}$. Hence, one may add a (massive propagating) scalar field  coupled to geometry; this leads to the {\sl scalar-tensor fourth order gravity}.

A fundamental point of studies is the possibility that ETG can be plagued by pathologies, such as the appearance of ghosts (negative norm states), which could allow for negative possibilities and consequently violation of unitarity~\cite{g1,g2,g3,g4}.  In particular, while standard GR and the Gauss-Bonnet theory have the same field content, this is not the case for the $f({\cal R})$ gravity type and Weyl gravity.
The former is safe, even though it does not improve the ultraviolet behaviour of the theory; $f({\cal R})$ gravity has just an extra scalar and can be ghost free in its Newtonian limit~\cite{ghosts1}. The latter has an extra pathological spin-2 field, which however causes no problem in the low-energy regime since then the effects of higher order terms give rise to small corrections to GR~\cite{ghosts1,ghosts2}. In our analysis, we will consider an action motivated from noncommutative geometry within the class of a Weyl type gravity. However, as we will discuss this proposal will be considered in the spirit of an effective field theory, hence is free from any pathologies.

Our starting point is the spherically symmetric solutions (in the newtonian limit) of scalar tensor fourth order gravity, then we study the dynamics of massless scalar field.

In this paper we try to obtain a new test-bed experiment, even if into a theoretical framework. We want to consider the Casimir effect in the curved space where the metric is given by a modified theory of gravity. Generally speaking, the Casimir effect \cite{casimir1,casimir2} can be defined as the stress on the bounding surfaces when a quantum field is confined in a finite volume of space. The confinement of physical field restricts obviously the modes of the corresponding quantum field giving us a measurable macroscopic effects.
The Casimir effect has been widley discussed in the flat space \cite{casimir_exp1,nesterenkoC,nesterenkoC2,nesterenkoC3,nesterenkoC4,nesterenkoC5,nesterenkoC6,nesterenkoC7,casimir_exp2,casimir_exp3, casimir_exp4}. They indeed have shown good agreement with the theory. Recently, some authors \cite{setare, calloni, caldwell,sorge} have considered the influence of gravitational field on the vacuum energy density of a quantum field inside a cavity. Indeed possible modifications in the vacuum energy induced by gravity could play a relevant role in the dynamics of the universe \cite{caldwell, brevik}. Also, at a microscopic level, Casimir energy modification in strong gravitational fields could become relevant in models of quark confinement based in string interquark potentials \cite{lambiase,lambiase2,lambiase3,lambiase4}. In connection with the equivalence principle there is a further question: whether vacuum fluctuations do gravitate or not. Finally, the analysis of the possible gravitational effects on Casimir cavities faces the open issue concerning the limits of validity of GR at small distances \cite{mostepanenko}. In this paper we perform a perturbative evaluation upto second order of the gravitational corrections to the  Casimir vacuum energy density for a massless scalar field confined in a cavity in a slightly curved static spacetime background.  Our starting point is the paper \cite{sorge} where the spacetime is given by the approximated Schwarzschild solution, while we consider the Newtonian limit of the solution in the case of ETG.

In what follows, we investigate in Sec.\ \ref{STFOG} the weak field limit of a particular ETG,  scalar-tensor-higher-order models, in view of constraining their parameters by analysis of Casimir effect. In Sec. \ref{casimireffect} we analyze the  dynamic of a massless scalar field in a weak gravitational field. Finally in Sec. \ref{constraints} we discuss the theoretical constraints of the ETG considered. Our conclusions are drawn in Sec.\ \ref{conclusions}

\section{The scalar tensor fourth order gravity}\label{STFOG}

Let us consider the action

\begin{eqnarray}\label{FOGaction}
\mathcal{S}\,=\,\int d^{4}x\sqrt{-g}\biggl[f({\cal R},{\cal R}_{\alpha\beta}{\cal R}^{\alpha\beta},\phi)+\omega(\phi)\phi_{;\alpha}\phi^{;\alpha}+\mathcal{X}\mathcal{L}_m\biggr],
\end{eqnarray}
where $f$ is an unspecified function of the Ricci scalar ${\cal R}$, the curvature invariant ${\cal R}_{\alpha\beta}{\cal R}^{\alpha\beta}\,\doteq\,Y$ where ${\cal R}_{\alpha\beta}$ is the Ricci tensor, and $\phi$ is a scalar field. We note that the Riemann tensor can be discarded since the Gauss-Bonnet invariant fixes it in the action (for details see Ref.~\cite{cqg}). Here $\mathcal{L}_m$ is the minimally coupled ordinary matter Lagrangian density, $\omega(\phi)$ is a generic function of the scalar field, $g$ is the determinant of metric tensor $g_{\mu\nu}$ and\footnote{Here we use the convention $c\,=\,1$.} $\mathcal{X}\,=\,8\pi G$.

In the metric approach, namely when the gravitational field is fully described by the metric tensor $g_{\mu\nu}$ only, the field equations are obtained by varying the action (\ref{FOGaction}) with respect to $g_{\mu\nu}$, leading to

\begin{eqnarray}\label{fieldequationFOG}
&&f_{\cal R}{\cal R}_{\mu\nu}-\frac{f+\omega(\phi)\phi_{;\alpha}\phi^{;\alpha}}{2}g_{\mu\nu}-f_{{\cal R};\mu\nu}+g_{\mu\nu}\Box
f_{\cal R}+2f_Y{{\cal R}_\mu}^\alpha
{\cal R}_{\alpha\nu}
\nonumber\\\\\nonumber&&-2[f_Y{{\cal R}^\alpha}_{(\mu}]_{;\nu)\alpha}+\Box[f_Y{\cal R}_{\mu\nu}]+[f_Y{\cal R}_{\alpha\beta}]^{;\alpha\beta}g_{\mu\nu}+\omega(\phi)\phi_{;\mu}\phi_{;\nu}\,=\,
\mathcal{X}\,T_{\mu\nu}~,
\end{eqnarray}
where $T_{\mu\nu}\,=\,-\frac{1}{\sqrt{-g}}\frac{\delta(\sqrt{-g}\mathcal{L}_m)}{\delta
g^{\mu\nu}}$ is the the energy-momentum tensor of matter, $f_{\cal R}\,=\,\frac{\partial f}{\partial {\cal R}}$, $f_Y\,=\,\frac{\partial f}{\partial Y}$ and $\Box={{}_{;\sigma}}^{;\sigma}$ is the D'Alembert operator. We use for the Ricci tensor the convention
${\cal R}_{\mu\nu}={{\cal R}^\sigma}_{\mu\sigma\nu}$, whilst for the Riemann
tensor we define ${{\cal R}^\alpha}_{\beta\mu\nu}=\Gamma^\alpha_{\beta\nu,\mu}+\cdots$. The
affinity connections are the usual Christoffel symbols of the metric, namely
$\Gamma^\mu_{\alpha\beta}=\frac{1}{2}g^{\mu\sigma}(g_{\alpha\sigma,\beta}+g_{\beta\sigma,\alpha}
-g_{\alpha\beta,\sigma})$, and we adopt the signature is $(+,-,-,-)$. The trace of the field equation
%(\ref{fieldequationFOG})
above, reads
\begin{eqnarray}\label{tracefieldequationFOG}
f_{\cal R}{\cal R}+2f_Y{\cal R}_{\alpha\beta}{\cal R}^{\alpha\beta}-2f+\Box[3
f_{\cal R}+f_Y{\cal R}]+2[f_Y{\cal R}^{\alpha\beta}]_{;\alpha\beta}-\omega(\phi)\phi_{;\alpha}\phi^{;\alpha}\,=\,\mathcal{X}\,T~,\end{eqnarray}
where $T\,=\,T^{\sigma}_{\,\,\,\,\,\sigma}$ is the trace of
energy-momentum tensor. Finally by varying the action (\ref{FOGaction}) with respect to the scalar field $\phi$,  we obtain the Klein-Gordon field equation
\begin{eqnarray}\label{FE_SF}
2\omega(\phi)\Box\phi+\omega_\phi(\phi)\phi_{;\alpha}\phi^{;\alpha}-f_\phi\,=\,0~,
\end{eqnarray}
where $\omega_\phi\,=\,\frac{d\omega}{d\phi}$ and $f_\phi\,=\,\frac{d f}{d\phi}$.

%%%%%%%%%%%%%%%%%%%%%%%%%%%%%%%%%%%%%%%%%%%%%%%%%%%%%%

\subsection{Solutions for a point-like source in the the weak field limit}

For many physical systems the study is computed in the weak-field approximation and in particular for our aim it is adequate the newtonian limit of the theory. In order to perform our approximation we have to perturb Eqs.~(\ref{fieldequationFOG}), (\ref{tracefieldequationFOG}) and (\ref{FE_SF}) in a Minkowski
background $\eta_{\mu\nu}$~\cite{PRD1,noi-newt}. This approximation is sometimes also referred to as an expansion in inverse powers of the speed of light. Neglecting the technical aspects we can set the expression of metric tensor $g_{\mu\nu}$ as follows

\begin{eqnarray}\label{PM_me}
&&g_{\mu\nu}\,\cong\,\begin{pmatrix}
1+g^{(2)}_{tt}(t,\mathbf{x})+\dots & 0 \\
0 & -\delta_{ij}+g^{(2)}_{ij}(t,\mathbf{x})+\dots\end{pmatrix}\,\doteq\,
\begin{pmatrix}
1+2\Phi & 0 \\
 & -\delta_{ij}+2\Psi\delta_{ij}\end{pmatrix}~,
\nonumber\\\\
&&\phi\,\cong\,\phi^{(0)}+\phi^{(2)}+\dots\,\doteq\,\phi^{(0)}+\varphi~,\nonumber
\end{eqnarray}
where $\Phi$, $\Psi$, $\varphi$ are proportional to the power $c^{-2}$ (Newtonian limit). The function $f$,  up to  the $c^{-4}$ order,  can be developed as

\begin{eqnarray}
\label{LimitFramework}
f({\cal R},{\cal R}_{\alpha\beta}{\cal R}^{\alpha\beta},\phi)\,=\,&&f_{\cal R}(0,0,\phi^{(0)})\,{\cal R}+\frac{f_{{\cal R}{\cal R}}(0,0,\phi^{(0)})}{2}\,{\cal R}^2+\frac{f_{\phi\phi}(0,0,\phi^{(0)})}{2}(\phi-\phi^{(0)})^2\nonumber\\\\\nonumber&&+f_{{\cal R}\phi}(0,0,\phi^{(0)}){\cal R}\,\phi+f_Y(0,0,\phi^{(0)}){\cal R}_{\alpha\beta}{\cal R}^{\alpha\beta}~,
\end{eqnarray}
while all other possible contributions in $f$ are negligible \cite{PRD1,mio2,FOGST}. By introducing the quantities

\begin{eqnarray}\label{mass_definition}
\begin{array}{ll}
{m_{\cal R}}^2\,\doteq\,-\frac{f_{\cal R}(0,0,\phi^{(0)})}{3f_{{\cal R}{\cal R}}(0,0,\phi^{(0)})+2f_Y(0,0,\phi^{(0)})}~,\\\\
{m_Y}^2\,\doteq\,\frac{f_{\cal R}(0,0,\phi^{(0)})}{f_Y(0,0,\phi^{(0)})}~,\\\\
{m_\phi}^2\,\doteq\,-\frac{f_{\phi\phi}(0,0,\phi^{(0)})}{2\omega(\phi^{(0)})}~,
\end{array}
\end{eqnarray}
and setting $f_{\cal R}(0,0,\phi^{(0)})\,=\,1$, $\omega(\phi^{(0)})\,=\,1/2$ for simplicity we get the complete set of differential equations

\begin{eqnarray}\label{PMfieldequationFOG3}
\begin{array}{ll}
(\triangle-{m_Y}^2)\triangle\Phi+\biggl[\frac{{m_Y}^2}{2}-\frac{{m_{\cal R}}^2+2{m_Y}^2}{6{m_{\cal R}}^2}\triangle\biggr]
{\cal R}+{m_Y}^2\,f_{{\cal R}\phi}(0,0,\phi^{(0)})\,\triangle\varphi\,=\,-{m_Y}^2\mathcal{X}\,\rho~,\\\\
\biggl\{(\triangle-{m_Y}^2)\triangle\Psi-\biggl[\frac{{m_Y}^2}{2}-\frac{{m_{\cal R}}^2+2{m_Y}^2}{6{m_{\cal R}}^2}\triangle\biggr]{\cal R}-{m_Y}^2\,f_{{\cal R}\phi}(0,0,\phi^{(0)})\,\triangle\varphi\biggr\}\delta_{ij}+\\\\\qquad+\biggl\{(\triangle-{m_Y}^2)(\Psi-\Phi)+\frac{{m_{\cal R}}^2-{m_Y}^2}{3{m_{\cal R}}^2}\,{\cal R}+{m_Y}^2\,f_{{\cal R}\phi}(0,0,\phi^{(0)})\,\varphi\biggr\}_{,ij}\,=\,0~,
\\\\
(\triangle-{m_{\cal R}}^2){\cal R}-3{m_{\cal R}}^2\,f_{{\cal R}\phi}(0,0,\phi^{(0)})\,\triangle\varphi\,=\,{m_{\cal R}}^2\,\mathcal{X}\,\rho~,\\\\
(\triangle-{m_\phi}^2)\varphi+f_{{\cal R}\phi}(0,0,\phi^{(0)})\,{\cal R}\,=\,0.
\end{array}
\end{eqnarray}
where also the energy momentum tensor $T_{\mu\nu}$ has been expanded in the case of perfect fluid when the pressure is negligible with respect to the mass density $\rho$. The development is started form the zeroth order, hence $T_{tt}\,=\,T^{(0)}_{tt}\,=\,\rho$ and $T_{ij}\,=\,T^{(0)}_{ij}\,=\,0$.

The last two equations of set (\ref{PMfieldequationFOG3}) are coupled system and, for a point-like source $\rho(\mathbf{x})\,=\,M\,\delta(\mathbf{x})$, admit the solutions

\begin{eqnarray}
\label{ST_FOG_FE_NL_sol_sys}
\begin{array}{ll}
\varphi(\textbf{x})\,=\,\sqrt{\frac{\xi}{3}}\,\frac{2 GM}{|\textbf{x}|}\frac{e^{-m_+\,|\textbf{x}|}-e^{-m_-\,|\textbf{x}|}}{\omega_+-\omega_-}~,\\\\
{\cal R}(\textbf{x})\,=\,-2{m_{\cal R}}^2\frac{ GM}{|\textbf{x}|}\frac{(\omega_+-\eta^2)\,e^{-m_+\,|\textbf{x}|}-(\omega_--\eta^2)\,e^{-m_-\,|\textbf{x}|}}{\omega_+-\omega_-}~,
\end{array}
\end{eqnarray}
where $\omega_\pm\,=\,\frac{1-\xi+\eta^2\pm\sqrt{(1-\xi+\eta^2)^2-4\eta^2}}{2}$, $m_\pm^2=m_{\cal R}^2\,\omega_\pm $, $\xi\,=\,3{f_{{\cal R}\phi}(0,0,\phi^{(0)})}^2$ and $\eta\,=\,\frac{m_\phi}{m_{\cal R}}$ \cite{FOGST}.

The solutions of the gravitational potential $\Phi$ and $\Psi$, derived from the first two  equations of set (\ref{PMfieldequationFOG3}) are

\begin{eqnarray}\label{ST_FOG_FE_NL_sol_point}
\Phi(\mathbf{x})\,=\,-\frac{GM}{|\mathbf{x}|}\biggl[1+g(\xi,\eta)\,e^{-m_+|\mathbf{x}|}+[\frac{1}{3}-g(\xi,\eta)]\,e^{-m_-|\mathbf{x}|}-\frac{4}{3}\,e^{-m_Y|\mathbf{x}|}\biggr]~,
\end{eqnarray}

\begin{eqnarray}\label{ST_FOG_FE_NL_sol_point_psi}
\Psi(\mathbf{x})\,=\,-\frac{GM}{|\mathbf{x}|}\biggl[1-g(\xi,\eta)\,e^{-m_+|\mathbf{x}|}-[1/3-g(\xi,\eta)]\,e^{-m_-|\mathbf{x}|}-\frac{2}{3}\,e^{-m_Y|\mathbf{x}|}\biggr]~,
\end{eqnarray}
where
 \[
 g(\xi,\eta)\,=\,\frac{1-\eta^2+\xi+\sqrt{\eta^4+(\xi-1)^2-2\eta^2(\xi+1)}}{6\sqrt{\eta^4+(\xi-1)^2-2\eta^2(\xi+1)}}\,~.
  \]
Note that for $f_Y\,\rightarrow\,0$ \emph{i.e.} $m_Y\,\rightarrow\,\infty$, we obtain the same outcome for the gravitational potential as in Ref.~\cite{FOGST} for a $f({\cal R},\phi)$-theory. The absence of the coupling term between the curvature invariant $Y$ and the scalar field $\phi$, as well as the linearity of the field equations (\ref{PMfieldequationFOG3}) guarantee that the solution (\ref{ST_FOG_FE_NL_sol_point}) is a linear combination of solutions obtained within an $f({\cal R},\phi)$-theory and an ${\cal R}+Y/{m_Y}^2$-theory generalizing the outcomes of the theory $f({\cal R},\,{\cal R}_{\alpha\beta}{\cal R}^{\alpha\beta})$ \cite{mio2}.

%%%%%%%%%%%%%%%%%%%%%%%%%%%%%%%%%%%%%%%%%%%%%%%%%%%%%%%%%%%%%%%

\section{Dynamic of a massless scalar field in a weak gravitational field}\label{casimireffect}

In a curved space-time the massless scalar field $\psi(t,\,\textbf{x})$ obeys the following field equation

\begin{eqnarray}\label{KG_Massless_Eq}
(\square +\xi \,{\cal R}) \psi(t,\,\textbf{x})\,=\,\frac{1}{\sqrt{-g}}\partial_{\alpha}\bigl[\sqrt{-g}\,g^{\alpha\beta}\,\partial_{\beta}\psi(t,\,\textbf{x})\bigr] + \xi \,{\cal R}\psi(t,\,\textbf{x})\,=\,0
\,\end{eqnarray}
where $\xi$ is a coupling parameter between geometry and matter. In our ETG framework, also in the vacuum, the curvature scalar ${\cal R}$ is different from zero (\ref{ST_FOG_FE_NL_sol_sys}). For simplicity we consider the massless scalar field $\psi(t,\,\textbf{x})$ confined between two parallel plates separated by a distance $L$ and with extension $S$, placed at a distance $R$ from the gravitational non-rotating source ($R\gg L, \sqrt{S}$), see figure.

\begin{figure}[htbp]
\centering
\includegraphics[scale=0.65]{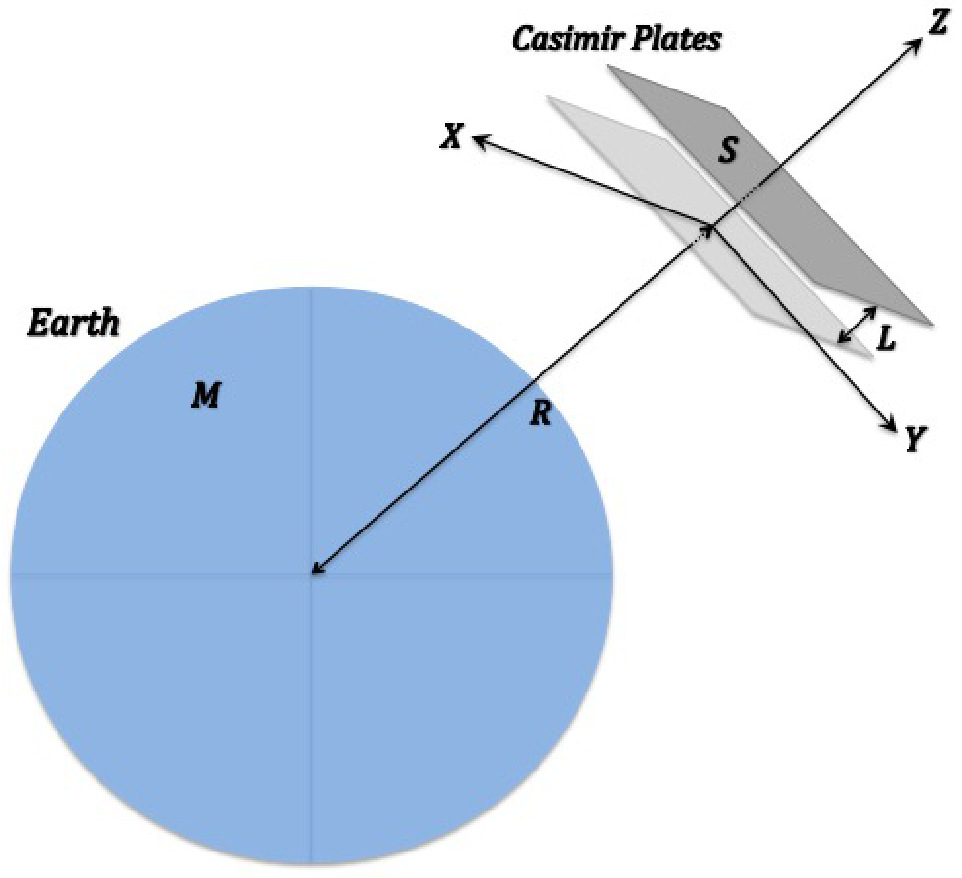}
%\caption{}
\label{fig_1}
\end{figure}

We choose a reference frame with the origin at one of the plates and the z-axis along the radial direction. We can expand the metric tensor components, using the gravitational potentials $\Phi(\textbf{x})$ (\ref{ST_FOG_FE_NL_sol_point}), $\Psi(\textbf{x})$ (\ref{ST_FOG_FE_NL_sol_point_psi}) and ${\cal R}(\textbf{x})$ (\ref{ST_FOG_FE_NL_sol_sys}), around the distance $R$ along the $z$ direction as follows

\begin{eqnarray}\label{Expantion_Potentials}
\begin{array}{ll}
g_{00}(\mathbf{x})\simeq  1+2\,\Phi_{0}(R) +2\,\Lambda(R)\,z
\\\\
g_{ij}(\mathbf{x})\simeq -1+2\,\Psi_{0}(R) +2\,\Sigma(R)\,z
\\\\
{\cal R}(\mathbf{x})\simeq {\cal R}_{1}(R)+{\cal R}_{2}(R)\,z
\end{array}
\end{eqnarray}
where

\begin{eqnarray}\label{Expantion_Potentials2}
\begin{array}{ll}
\Phi_{0}(R)\,=\,-\frac{GM}{R}\biggl[1+g(\xi,\eta)\,e^{-m_+ R}+\biggl(\frac{1}{3}-g(\xi,\eta)\biggr)\,e^{-m_- R}-\frac{4}{3}\,e^{-m_Y R}\biggr]
\\\\
\Lambda(R)\,=\,\frac{GM}{R^2}\biggl[1+g(\xi,\eta)\ \bigl(1+m_+ R\bigl)\,e^{-m_+ R}+\biggl(\frac{1}{3}-g(\xi,\eta)\biggr)\bigl(1+m_- R\bigl) \,e^{-m_- R}+\\\\
\qquad\qquad\qquad\qquad\qquad\qquad\qquad\qquad\qquad\qquad\qquad\qquad\qquad-\frac{4}{3}\bigl(1+m_Y R\bigl) \,e^{-m_Y R}\biggr]
\\\\
\Psi_{0}(R)\,=\,-\frac{GM}{R}\biggl[1-g(\xi,\eta)\,e^{-m_+ R}-\biggl(\frac{1}{3}-g(\xi,\eta)\biggr)\,e^{-m_- R}-\frac{2}{3}\,e^{-m_Y R}\biggr]
\\\\
\Sigma(R)\,=\,\frac{GM}{R^2}\biggl[1-g(\xi,\eta)\bigl(1+m_+ R\bigl)\,e^{-m_+ R}-\biggl(\frac{1}{3}-g(\xi,\eta)\biggr)\bigl(1+m_- R\bigl)\,e^{-m_- R}+\\\\
\qquad\qquad\qquad\qquad\qquad\qquad\qquad\qquad\qquad\qquad\qquad\qquad\qquad-\frac{2}{3}\bigl(1+m_Y R\bigl)\,e^{-m_Y R}\biggr]
\\\\
{\cal R}_{1}(R)\,=\,-{2 m_{\cal R}}^2\frac{GM}{R}\frac{(\omega_+-\eta^2)\,e^{-m_+ R}-(\omega_--\eta^2)\,e^{-m_- R}}{\omega_+-\omega_-}
\\\\
{\cal R}_{2}(R)\,=\,{2 m_{\cal R}}^2\frac{GM}{R^2}\frac{(\omega_+-\eta^2)\,\bigl(1+m_+ R\bigl)e^{-m_+ R}-(\omega_--\eta^2)\bigl(1+m_- R\bigl)\,e^{-m_- R}}{\omega_+-\omega_-}
\end{array}
\end{eqnarray}

Using the metric (\ref{Expantion_Potentials}), the field equation for the scalar field $\psi(t,\,\textbf{x})$ becomes

\begin{eqnarray}\label{KG_Massless_Eq_Expantion}
{\ddot \psi}(t,\,\textbf{x})-\biggl[1+4\Phi+ 4\gamma z \biggr] \vartriangle \psi(t,\,\textbf{x})+\xi\,\biggl[{\cal R}_{1}+{\cal R}_{2} z\biggl] \psi(t,\,\textbf{x})=\,0
\end{eqnarray}
where

\begin{eqnarray}
\begin{array}{ll}
\Phi\equiv\Phi_{0}(R)+\Psi_{0}(R)\,=\,-\frac{2GM}{R}\biggl[1-\,e^{-m_Y R}\biggr]
\\\\
\gamma\equiv\Lambda(R)+\Sigma(R)\,=\,\frac{2GM}{R^2}\biggl[1-\bigl(1+m_Y R\bigl) \,e^{-m_Y R}\biggr]
\end{array}
\end{eqnarray}

Let us solve the equation (\ref{KG_Massless_Eq_Expantion}) and we suppose the solution of the form

\begin{eqnarray}\label{Ciccio}
\psi(t,\,\textbf{x})\,=\,N\,\Xi(z)\,e^{i\,(\omega t-\textbf{k}_\bot\cdot\textbf{x}_\perp)}
\end{eqnarray}
where $\textbf{k}_\bot\equiv(k_x,k_y)$, $\textbf{x}_\bot\equiv(x,y)$ and $N$ is a normalization constant. The field equation (\ref{KG_Massless_Eq_Expantion}) becomes

\begin{eqnarray}\label{KG_Massless_Eq_Expantion2}
(\partial^2_\lambda+\lambda)\,\Xi(\lambda)\,=\,0
\end{eqnarray}
where
\begin{eqnarray}\label{theta_zita}
\begin{array}{ll}
\lambda\equiv\lambda(z)\,=\,\alpha-\beta\,z\equiv\,\zeta\,\theta^{-2/3}-\theta^{1/3}\,z
\\\\
\zeta\,=\,\bigl( 1-4\Phi\bigr)\omega^2-\textbf{k}_\bot^2-\xi\,{\cal R}_1
\\\\
\theta\,=\,4\omega^2\,\gamma+\xi\,{\cal R}_2

\end{array}
\end{eqnarray}
The solution of (\ref{KG_Massless_Eq_Expantion2}) can be given by means Bessel functions:
\begin{eqnarray}\label{Bessel_Equation}
\Xi(\lambda)\,=\,\sqrt{\lambda}\biggl[ C_1\,J_{1/3}\biggl( \frac{2}{3}\lambda^{3/2}\biggr)+C_2\,J_{-1/3}\biggl( \frac{2}{3}\lambda^{3/2}\biggr) \biggr]\,,
\end{eqnarray}
where $C_1$ and $C_2$ are constants. We note that $\lambda\gg 1$ 	because $\theta\ll\zeta$. Then we can use the asymptotic form of (\ref{Bessel_Equation}):
\begin{eqnarray}\label{Bessel_Equation_asymptotic}
\Xi(\lambda)\,\simeq\,\sqrt{\frac{3}{\pi\,\sqrt{\lambda}}}\sin\biggl[ \frac{2}{3}\lambda^{3/2}+\tau \biggr]\,.
\end{eqnarray}
If we assume that the field $\psi$ satisfies Dirichlet boundary conditions on the plates, that is:
\begin{eqnarray}\label{Dirichlet_boundary_conditions}
\psi(z=0)\,=\,\psi(z=L)\,=\,0 \,,
\end{eqnarray}
after some algebra, we find the relation:
\begin{eqnarray}\label{Dirichlet_Relation}
\frac{2}{3}\biggl[ \lambda^{3/2}\bigl( 0\bigr)-\lambda^{3/2}\bigl( L\bigr)\biggr]\,=\,n\,\pi,
\end{eqnarray}
with $n$ is integer number. By the relation above (\ref{Dirichlet_Relation}) we find the energy spectrum:
\begin{eqnarray}\label{Energy_Spectrum}
\omega^2_n\,=\,\bigl(1+4\Phi+2\gamma\,L\bigr)\biggl[\textbf{k}_\bot^2+\biggl( \frac{n\pi}{L}\biggr)^2\biggr]+\xi\,\biggl[ {\cal R}_1+\frac{1}{2}{\cal R}_2\,L\biggr]\,,
\end{eqnarray}
We note that in the energy spectrum the dependence on parameters $m_{\cal R}$ and $m_\phi$ is only within the terms related to the curvature, the second term in the right-hand side equation (\ref{Energy_Spectrum}). In other word, if the coupling parameter $\xi$ were zero the energy spectrum in the case of $f({\cal R},\phi)$ gravity we will be the same of GR.

Finally, using the scalar product defined in the theory of quantum field in curved spacetimes:
\begin{eqnarray}\label{scalar_product}
\bigl(\psi_n,\psi_m\bigl)\,=\,-i\int_V\biggl[\bigl(\partial_{\mu}\,\psi_n\bigr)\psi^*_m-\psi_n\bigl(\partial_{\mu}\,\psi_m\bigr)^*\biggr]\sqrt{-g_3}\,n^\mu\, dx\,dy\,dz\,,
\end{eqnarray}
we find the normalization constant:
\begin{eqnarray}\label{N^2}
N_n^2\,=\,\frac{\alpha\,\beta}{3\,S\,\omega_n\,n\bigl( 1-\Phi_0\bigl)}.
\end{eqnarray}
\subsection{Casimir vacuum density energy}

To calculate the mean vacuum energy density ${\cal E}$ between the plates, we use the general relation:
\begin{eqnarray}
 {\cal E}\,=\,\frac{1}{V_P}\sum_n\int d^2\,\textbf{k}_\bot\int dx\,dy\,dz\,\sqrt{-g_3} \,\bigl( g_{00}\bigr)^{-1}\,T_{00}\bigl( \psi_n,\psi_n^*\bigr),
\end{eqnarray}
where
\begin{eqnarray}\label{Proper_Volume_Energy_Momentum_Tensor}
&&L_P\,=\,\int dz\,\sqrt{-g_3}\,\simeq \, L\,\biggl[1-\Psi_0-\frac{1}{2} \Sigma\,L\biggl]\\
\nonumber
&&V_P\,=\,\int dx\,dy\,dz\,\sqrt{-g_3}\,\simeq \, S\,L\biggl[1-3\Psi_0-\frac{3}{2}\Sigma L\biggl]\\
\nonumber
&&T_{\mu\nu}\,=\,\partial_{\mu}\,\psi\,\partial_{\nu}\,\psi-\frac{1}{2}g_{\mu\nu}g^{\alpha\beta}\partial_{\alpha}\,\psi\,\partial_{\beta}\,\psi\,,
\end{eqnarray}
$L_P$ is the proper length of the cavity, $V_P$ is the the proper volume and $T_{00}\bigl( \psi_n,\psi_n^*\bigr)$ is a component of energy-momentum tensor.
Using Schwinger proper-time representation and $\zeta$-function regularization, we find the mean vacuum energy density:
\begin{eqnarray}\label{Energy_Casimir_2_Order}
{\cal E}\,=\,-\biggl[1+3\bigl(\Phi_0-\Psi_0\bigr) -\bigl( 2\Sigma-\Lambda\bigr) L_P\biggr]\frac{\pi^2}{1440\,L^4_P}+\frac{\xi\,{\cal R}_2}{192\,L_P}\,,
\end{eqnarray}
where,
\begin{eqnarray}
\nonumber
&&\Phi_0-\Psi_0\,=\,-\frac{2GM}{R}\biggl[g(\xi,\eta)\,e^{-m_+ R}+\biggl(\frac{1}{3}-g(\xi,\eta)\biggr)\,e^{-m_- R}-\frac{1}{3}\,e^{-m_Y R}\biggr]\\
\nonumber
%&&\simeq GM\biggl[g(\xi,\eta)\,m_+(2-m_+ R)+\biggl(\frac{1}{3}-g(\xi,\eta)\biggr)\,\,m_-(2-m_- R)-\frac{1}{3}\,\,m_Y(2-m_Y R)\biggr]\\
%\nonumber
&& 2\Sigma-\Lambda\,=\,\frac{GM}{R^2}\biggl[1-3\,g(\xi,\eta)\bigl(1+m_+ R\bigl)\,e^{-m_+ R}-
\biggl(1-3g(\xi,\eta)\biggr)\bigl(1+m_- R\bigl)\,e^{-m_- R}\biggr]\,.
\end{eqnarray}
The relation (\ref{Energy_Casimir_2_Order}) gives us the corrections of the Casimir vacuum energy density, up to the order $o(M/R^2)$ \footnote{We have neglected in the (\ref{Energy_Casimir_2_Order}) all products of order higher than $c^{-2}$.}, for the ETG (\ref{FOGaction}) in the weak field limit. The GR give us  corrections to Casimir vacuum energy density only at second order o$(M/R^2)$ \cite{sorge}, but in the ETG we find also corrections of the fist order $o(M/R)$. Therefore, to the Casimir density energy the ETG gives a greater contribution than those of GR. This is a very interesting result, because we found a physical quantity testing directly ETG. Finally, we note that the correction of fist order increases the Casimir vacuum energy density, while those of the second order decreases the Casimir vacuum energy density.

\subsection{Casimir pressure}

The attractive force observed between the cavity plates is obtained by the relation ${\cal F}=-\frac{\partial E}{\partial L_P}$, where $E={\cal E}\,V_P$ is the Casimir vacuum energy, in our case we get:

\begin{eqnarray}\label{Force_Casimir}
{\cal F}\,=\,-\biggl[1+3\bigl(\Phi_0-2\Psi_0\bigr) -\frac{7\Sigma-2\Lambda}{3} L_P\biggr]\frac{\pi^2\,S_P}{480\,L^4_P}\,,
\end{eqnarray}

where

\begin{eqnarray}
\nonumber
&&\Phi_0-2\Psi_0\,=\,\frac{GM}{R}\biggl[1-3g(\xi,\eta)\,e^{-m_+ R}-\bigl(1-3g(\xi,\eta)\bigr)\,e^{-m_- R}\biggr]\\
\nonumber
&& 7\Sigma-2\Lambda\,=\,\frac{GM}{R^2}\biggl[5+9\,g(\xi,\eta)\bigl(1+m_+ R\bigl)\,e^{-m_+ R}+3
\bigl(1-3g(\xi,\eta)\bigr)\bigl(1+m_- R\bigl)\,e^{-m_- R}+\\
\nonumber
&&\qquad\qquad\qquad\qquad\qquad\qquad\qquad\qquad\qquad\qquad\qquad\qquad\qquad\qquad-8\bigl(1+m_Y R\bigl)\,e^{-m_Y R}\biggr]\,.
\end{eqnarray}
We note that the the contributions of curvature in the force (\ref{Force_Casimir}) are zero mathematically. Furthermore, the correction at first order does not depend on the parameter $m_Y$, while  those of the second order depend on all. For this reason, to obtain corrections at first order to the GR results the more general ETG (\ref{FOGaction}) is simply $f({\cal R},\phi)$.

In the end, we introduce the Casimir pressure ${\cal P}={\cal F}/S$,
\begin{eqnarray}\label{Pressure_Casimir}
{\cal P}\,&=&\,{\cal P}_0+{\cal P}_{ETG}\,,\\
\nonumber
{\cal P}_0\,&=&\,-\frac{\pi^2\,}{480\,L^4_P}\,,\\
\nonumber
{\cal P}_{ETG}\,&=&\,\biggl[3\bigl(\Phi_0-2\Psi_0\bigr) -\frac{7\Sigma-2\Lambda}{3} L_P\biggr]{\cal P}_0\,,
\end{eqnarray}
where ${\cal P}_0$ is the Casimir pressure in the flat case, while ${\cal P}_{ETG}$ is the correction to the pressure in the context of ETG.

\section{Experimental constraints}\label{constraints}

One has to test the observational compatibility of an ETG. To do this, we can use the pressure as a measurable physical quantity. Imposing the constraint $|{\cal P}_{ETG}|\lesssim \delta {\cal P}$, where $\delta {\cal P}$ is the experimental error, we obtain the following relation
\begin{eqnarray}\label{Contraint_Pressure_Casimir}
\biggl|3\bigl(\Phi_0-2\Psi_0\bigr) -\frac{7\Sigma-2\Lambda}{3} L_P\biggr|\lesssim \frac{\delta {\cal P}}{{\cal P}_0}\,,
\end{eqnarray}
that gives us the the constraint for the free parameters (\ref{mass_definition}) of the ETG.

Now, we analyze some models of ETG, see Table \ref{tab:tab2}.
\begin{table}[!ht]
\centering
\begin{tabular}{c|c|c}
\hline\hline\hline
Case & ETG & Mass definition \\
\hline
& & \\
A & $f({\cal R})$ & $\begin{array}{ll}{m_{\cal R}}^2\,=\,-\frac{1}{3f_{{\cal R}{\cal R}}(0)}\\\\{m_Y}^2\,\rightarrow\,\infty,\,\,\,{m_\phi}^2\,=\,0\\\\\xi\,=\,0,\,\,\,\eta\,=\,0,\,\,\,g(\xi,\eta)=2/3\\\\m_+\,=\,m_{\cal R},\,\,\,m_-\,\rightarrow\,\infty
\end{array}$  \\
\hline
& & \\
B & $f({\cal R},\,{\cal R}_{\alpha\beta}{\cal R}^{\alpha\beta})$ & $\begin{array}{ll}{m_{\cal R}}^2\,=\,-\frac{1}{3f_{{\cal R}{\cal R}}(0,\,0)+2f_Y(0,\,0)}\\\\{m_Y}^2\,=\,\frac{1}{f_Y(0,\,0)},\,\,\,{m_\phi}^2\,=\,0\\\\\xi\,=\,0,\,\,\,\eta\,=\,0,\,\,\,g(\xi,\eta)=2/3\\\\m_+\,=\,m_{\cal R},\,\,\,m_-\,\rightarrow\,\infty
\end{array}$  \\
\hline
&  & \\
C & $f({\cal R},\,\phi)+\omega(\phi)\phi_{;\alpha}\phi^{;\alpha}$ & $\begin{array}{ll}{m_{\cal R}}^2\,=\,-\frac{1}{3f_{{\cal R}{\cal R}}(0,\,\phi^{(0)})}\\\\{m_Y}^2\,\rightarrow\,\infty,\,\,\,\,{m_\phi}^2\,=\,-f_{\phi\phi}(0,\,\phi^{(0)})\\\\\xi\,=\,3{f_{{\cal R}\phi}(0,\,\phi^{(0)})}^2,\,\,\,\,\eta\,=\,\frac{m_\phi}{m_{\cal R}}\\\\m_\pm\,=\,\sqrt{\frac{1-\xi+\eta^2\pm\sqrt{(1-\xi+\eta^2)^2-4\eta^2}}{2}}\,m_{\cal R}
\end{array}$  \\
\hline
&  & \\
D & $c_0{\cal R}+c_1\,{\cal R}\,\phi+f(\phi)+\omega(\phi)\phi_{;\alpha}\phi^{;\alpha}$ & $\begin{array}{ll}{m_{\cal R}}^2\rightarrow\,\infty\,\,,\,{m_Y}^2\rightarrow\,\infty,\,\,\,\,{m_\phi}^2\,=\,-f_{\phi\phi}(\phi^{(0)})\\\\\xi\,=\,3\,{c_1}^2,\,\,\,\,\eta\,\rightarrow\,0,\,\,\,g(\xi,0)=\frac{2}{3(1-\xi)}\\\\m_+\,\rightarrow\,\infty,\,\,\,m_-\,\rightarrow\,\frac{m_\phi}{\sqrt{1-3\,{c_1}^2}}
\end{array}$  \\
\hline
&  & \\
E & $f({\cal R},\,{\cal R}_{\alpha\beta}{\cal R}^{\alpha\beta},\phi)+\omega(\phi)\phi_{;\alpha}\phi^{;\alpha}$ & $\begin{array}{ll}{m_{\cal R}}^2\,=\,-\frac{1}{3f_{{\cal R}{\cal R}}(0,\,0,\,\phi^{(0)})+2f_Y(0,\,0,\,\phi^{(0)})}\\\\{m_Y}^2\,=\,\frac{1}{f_Y(0,0,\phi^{(0)})},\,\,\,\,{m_\phi}^2\,=\,-f_{\phi\phi}(0,\,0,\,\phi^{(0)})\\\\\xi\,=\,3{f_{{\cal R}\phi}(0,0,\phi^{(0)})}^2,\,\,\,\,\eta\,=\,\frac{m_\phi}{m_{\cal R}}\\\\m_\pm\,=\,\sqrt{\frac{1-\xi+\eta^2\pm\sqrt{(1-\xi+\eta^2)^2-4\eta^2}}{2}}\,m_{\cal R}
\end{array}$  \\
\hline\hline\hline
\end{tabular}
\caption{\label{tab:tab2}Here $f_{\cal R}(0,\,0,\,\phi^{(0)})\,=\,1$ and $\omega(\phi^{(0)})\,=\,1/2$ and for the case D we set also $c_0+c_1\phi^{(0)}\,=\,1$.}
\end{table}
Let us consider case A of Table; the only interesting quantity is $m_{\cal R}$. For the relation (\ref{Contraint_Pressure_Casimir}) implies
\begin{eqnarray}
\label{FR_constraint}
\biggl| 1-2\,e^{-m_{\cal R} R_\oplus}\biggr|\lesssim \frac{2}{3}\frac{R_\oplus}{R_\oplus^S}\frac{\delta {\cal P}}{{\cal P}_0}\,,
\end{eqnarray}
where $R_\oplus$ and $R_\oplus^S$ are the radius and Schwarzschild radius of the Earth. As a special case of $f({\cal R})$-theories one can consider the polynomial expression

\begin{eqnarray}
f({\cal R})\,=\,{\cal R}+\alpha\,{\cal R}^2+\sum_{n\,=\,3}^{N}\alpha_n\,{\cal R}^n\,.
\end{eqnarray}
Note however that the characteristic scale $m_{\cal R}$ is only generated by the ${\cal R}^2$-term. In literature an interesting model of $f({\cal R})$-theories is that of Starobinsky $f({\cal R})\,=\,{\cal R}-{\cal R}^2/{\cal R}_0$~\cite{staro}, for which $m_{\cal R}^2={\cal R}_0/6$.

To generalize the previously result we must include the  curvature invariant ${\cal R}_{\mu\nu}{\cal R}^{\mu\nu}$. For case B of Table~\ref{tab:tab2} we consider the general class of $f({\cal R},\,{\cal R}_{\alpha\beta}{\cal R}^{\alpha\beta})$-theories and their characteristic scales $m_{\cal R}$ and $m_Y$. Using equation (\ref{Contraint_Pressure_Casimir}), we obtain
 \begin{eqnarray}
\nonumber
\biggl| 1-2\,e^{-m_{\cal R} R_\oplus}-\frac{1}{6}\frac{L_P}{R_\oplus}\biggr[5+6\bigl(1+m_{\cal R} R_\oplus\bigl)\,e^{-m_{\cal R} R_\oplus}-8\bigl(1+m_Y R_\oplus\bigl)\,e^{-m_Y R_\oplus}\biggl]\biggl|\lesssim \frac{2}{3}\frac{R_\oplus}{R_\oplus^S}\frac{\delta {\cal P}}{{\cal P}_0}\,.
\end{eqnarray}
We note that to constrain also the parameter $m_Y$ we need of the corrections at second order. This class of theories includes the case of a Weyl square type model, \emph{i.e.} $C_{\mu\nu\rho\sigma}C^{\mu\nu\rho\sigma}\,=\,2{\cal R}_{\mu\nu}{\cal R}^{\mu\nu}-\frac{2}{3}{\cal R}^2$, where there is only one characteristic scale $m_{\cal R}\,\rightarrow\,\infty$.

The same argumentation is also valid for the scalar-tensor case of theory, for which in the Newtonian limit (see case D in Table \ref{tab:tab2}) the more general expression (\ref{LimitFramework}) becomes

\begin{eqnarray}\label{4.4}
\biggl(1-\phi^{(0)}\sqrt{\frac{\xi}{3}}\biggr){\cal R}+\sqrt{\frac{\xi}{3}}\,{\cal R}\,\phi-\frac{{m_\phi}^2}{2}\,(\phi-\phi^{(0)})^2~.
\end{eqnarray}
Thus, for the most general Scalar-Tensor (ST) theory in the Newtonian limit, one can consider the model\footnote{With the condition $\alpha_0+\alpha_1\,\phi^{(0)}\,=\,1$.}

\begin{eqnarray}\label{scalar-tensor}
f_{\rm ST}({\cal R},\phi)\,=\,c_0\,{\cal R}+c_1\,{\cal R}\,\phi-\frac{1}{2}{m_\phi}^2\,(\phi-\phi^{(0)})^2+\frac{1}{2}\phi_{,\alpha}\phi^{,\alpha}~.
\end{eqnarray}
Since for this case $m_{\cal R}\rightarrow \infty $, $m_Y\rightarrow \infty$, $\xi\,=\,3\,{c_1}^2$, $\eta\,\rightarrow\,0$, $m_+\rightarrow \infty$ and  $m_-\,=\,\frac{m_\phi}{\sqrt{1-3{c_1}^2}}$, we obtain from equation (\ref{Contraint_Pressure_Casimir})

\begin{eqnarray}
\nonumber
\biggl| 1-\frac{2\,e^{-m_- R_\oplus}}{1-3c_1^2}\biggr|\,\lesssim \frac{2}{3}\frac{R_\oplus}{R_\oplus^S}\frac{\delta {\cal P}}{{\cal P}_0}\,,
\end{eqnarray}
As a special case of  a scalar-tensor fourth order gravity model (case E) we consider NonCommutative Spectral Geometry (NCSG)~\cite{connes_1,connes_2}, for which at a cutoff scale (set as the Grand Unification scale) the purely gravitational part of the action coupled to the Higgs ${\bf H}$ reads~\cite{ccm}
\begin{eqnarray}\label{eq:action1}
{\cal S}_{\rm NCSG}\,&=&\,\int d^4x\sqrt{-g}\biggl[\frac{{\cal R}}{2{\kappa_0}^2}+ \alpha_0\,C_{\mu\nu\rho\sigma}C^{\mu\nu\rho\sigma}
+\tau_0 {\cal R}^\star {\cal R}^\star+\frac{{\bf H}_{;\alpha}{\bf H}^{;\alpha}}{2}-{\mu_0}^2\,
{\bf H}^2+\\
\nonumber
&&\,\,\,\,\,\,\,\,\,\,\,\,\,\,\,\,\,\,\,\,\,\,\,\,\,\,\,\,\,\,\,\,\,\,\,\,\,\,\,\,\,\,\,\,\,\,\,\,\,\,\,\,\,\,\,\,\,\,\,\,\,\,\,\,\,\,\,\,\,\,\,\,\,\,\,\,\,\,\,\,\,\,\,\,\,\,\,\,\,\,\,\,\,\,\,\,\,\,\,\,\,\,\,\,\,\,\,\,\,\,\,\,\,\,\,\,\,\,\,\,\,\,\,\,\,\,\,\,\,\,\,\,\,\,\,\,-\frac{{\cal R}\,{\bf H}^2}{12}+\lambda_0\,{\bf H}^4\biggr]\,,
\end{eqnarray}
where ${\cal R}^\star {\cal R}^\star$ is the topological term that integrates to the Euler characteristic, hence nondynamical. Since the square of the Weyl tensor can be expressed in terms of ${\cal R}^2$ and ${\cal R}_{\mu\nu}{\cal R}^{\mu\nu}$ as $C_{\mu\nu\rho\sigma}C^{\mu\nu\rho\sigma}\,=\,2{\cal R}_{\mu\nu}R^{\mu\nu}-\frac{2}{3}{\cal R}^2$, the NCSG action is a particular case of action (\ref{FOGaction}). For this model, we have the following parameters

\begin{eqnarray}
\begin{array}{ll}
m_R\rightarrow \infty\qquad\qquad\qquad\qquad\qquad\qquad \xi=\frac{af_0{\bf H}_{0}^2}{12\pi^2}
\\\\
m_Y=\sqrt{\frac{5\pi^2\bigl(k_0^2{\bf H}_{0}-6\bigr)}{36f_0k_0^2}}\qquad\qquad\qquad\,\,\,\,\,\eta\,\rightarrow 0
\\\\
m_\phi=\frac{a\,f_0}{\pi^2}(2\mu_0-12\lambda_0\,{\bf H}_{0}^2)\qquad\qquad\,\,\,\,\,m_+\rightarrow \infty,\,\,\,m_-\rightarrow \frac{m_\phi}{\sqrt{1-\xi}}
\end{array}
\end{eqnarray}
if we set $f_R(0,\,0,\,\phi^{(0)})\,=\,\frac{1}{2{k_0}^2}-\xi_0\frac{a\,f_0}{\pi^2}\,{\phi^{(0)}}^2\,=\,1$. Using the equation (\ref{Contraint_Pressure_Casimir}) we obtain
 \begin{eqnarray}
\nonumber
\biggl| 1-\frac{1}{6}\frac{L_P}{R_\oplus}\biggr[5-8\bigl(1+m_Y R_\oplus\bigl)\,e^{-m_Y R_\oplus}\biggl]\biggl|\lesssim \frac{2}{3}\frac{R_\oplus}{R_\oplus^S}\frac{\delta {\cal P}}{{\cal P}_0}\,.
\end{eqnarray}

The total absolute experimental error in the experiment on measuring the Casimir pressure between Au coated plates by means of micro-mechanical torsional oscillator at shortest separations is as small as 0.2\% ($ \delta {\cal P}/{\cal P}_0\simeq 0.002$) of the measured Casimir pressure \cite{mostepanenko2}. Hence, today the current technology is far from allowing a direct experiment check of the influence of gravity on Casimir pressure\footnote{We note that for the Earth $\Phi_0-2\Psi_0 \simeq 10^{-9}$ and $(7\Sigma-2\Lambda)L_P \simeq 10^{-22}$.}.

In fact, for sake of simplicity we can analyze the case of a $f({\cal R})$-theory (\ref{FR_constraint}). The term on the right side hand of (\ref{FR_constraint}) is almost $10^6$ \footnote{ We have $\frac{R_\oplus}{R_\oplus^S}\simeq 10^9$ and $\frac{\delta {\cal P}}{{\cal P}_0}\simeq 10^{-3}$.}, while the one on the left side hand is proportional to unity. For this reason, today, we can not use the experiment values on measuring the Casimir pressure \cite{mostepanenko2} to constrain the parameter $m_{{\cal R}}$, but we need to improve at least of six orders of magnitude the experimental sensitivity on Heart, $\frac{\delta {\cal P}}{{\cal P}_0}\lesssim 10^{-9}$, to be able to constrain the free parameter. However, today there is a big interest in the gravitational interferometers, since they have reached a high sensitivity,  and then they could in the future be used as the tool to detect these small effect induced by gravity on Casimir pressure and therefore improve the six orders of magnitude.

Besides, future space missions, with advanced technology, might make some experimental measurements around the planet Jupiter. Indeed, in this case, we have the best ratio of the radii, $\frac{R_J}{R_J^S}\simeq 10^7$, in all the solar system and then we should only improve four orders of magnitude in the instrument sensitivity. Hence, in the distant future Jupiter could be a different laboratory in the Solar System to check the gravity influence on Casimir pressure.

\section{Conclusions}\label{conclusions}

Working in the weak field approximation, we have solved the field equation in a curved space-time for a massless scalar field, following a perturbative approach up to the o$(M/R^2)$ order. We have derived the correction to the flat spacetime  Casimir vacuum energy density and pressure in the context of ETG.

For the Casimir vacuum energy density (\ref{Energy_Casimir_2_Order}) we found that GR does not give contribution at fist order, but only at second one. This is a very interesting result because future experiments with higher sensitivity may be tested directly  the ETG.

Lastly, in the equations (\ref{Pressure_Casimir}) we found the relation for the Casimir pressure. In this case GR gives contribution also at first order. Requiring that the correction were within the experimental errors, we then imposed constraints (\ref{Contraint_Pressure_Casimir}) on the free parameters for a given ETG. However, today we can not use the data available to us \cite{mostepanenko2} to constrain the free parameters, because the sensitivity of the experiments need to improve at least six order of magnitude to quantify the small effect induced by gravity. For these reason, the gravitational interferometers, with the achieved high sensitivity, could give a direct check of the influence of gravity on Casimir pressure and then be an alternative tool to test the ETG. On the other hand, we discussed how the planet Jupiter could be a good laboratory, in the solar system, to test some ETG measuring the Casimir pressure with appropriate space missions around it.

\end{document}